\pdfoutput=1

\documentclass[twocolumn,english,superscriptaddress]{revtex4}
\usepackage{times,amssymb,amsmath,graphicx}
\usepackage[bookmarks=false,pdffitwindow=false,pdfstartview={FitH}]{hyperref}
\usepackage[T1]{fontenc}
\usepackage{babel}
\usepackage{color}

\begin{document}


\title{
  Mott transition in the Hubbard model on anisotropic honeycomb lattice with
  implications for strained graphene: Gutzwiller variational study
}

\author{Grzegorz Rut}
\affiliation{Institute for Theoretical Physics,
  Jagiellonian University, \L{}ojasiewicza 11, PL--30348 Krak\'{o}w, Poland}
\affiliation{Verisk Analytics Sp.\ z~o.o., 
  Rakowicka 7, PL--31511 Krak\'{o}w, Poland}

\author{Maciej Fidrysiak}
\affiliation{Institute for Theoretical Physics,
  Jagiellonian University, \L{}ojasiewicza 11, PL--30348 Krak\'{o}w, Poland}

\author{Danuta Goc-Jag\l{}o}
\affiliation{Institute for Theoretical Physics,
  Jagiellonian University, \L{}ojasiewicza 11, PL--30348 Krak\'{o}w, Poland}

\author{Adam Rycerz}
\affiliation{Institute for Theoretical Physics,
  Jagiellonian University, \L{}ojasiewicza 11, PL--30348 Krak\'{o}w, Poland}

\date{November 30, 2022}

\begin{abstract}
Modification of interatomic distances due to high pressure leads to exotic phenomena, including metallicity, superconductivity and magnetism, observed in materials not showing such properties in normal conditions. In two-dimensional crystals, such as graphene, atomic bond lengths can be modified by more that 10 percent by applying in-plane strain, i.e., without generating high pressure in the bulk. In this work, we study the strain-induced Mott transition on a~honeycomb lattice by using computationally inexpensive techniques, including Gutzwiller Wave Function (GWF) and different variants of Gutzwiller Approximation (GA), obtaining the lower and upper bounds for critical Hubbard repulsion ($U$) of electrons. For uniaxial strain in the armchair direction the band gap is absent, and electron correlations play a~dominant role. A significant reduction of the critical Hubbard $U$ is predicted. Model considerations are mapped onto tight-binding Hamiltonian for monolayer graphene by the auxiliary Su-Schrieffer-Heeger model for acoustic phonons, assuming zero stress in the direction perpendicular to the strain applied. Our results suggest that graphene, although staying in semimetallic phase even for extremely high uniaxial strains, may show measurable signatures of electron correlations, such as the band narrowing and the reduction of double occupancies. 
\end{abstract}

\maketitle


\section{Introduction}
The Hubbard model, initially proposed to describe interaction-driven
transition between conducting and insulating systems \cite{Gut63,Hub63},
needs to be carefully applied in low dimensions, where exact solutions
(when available) \cite{Lie68,Lie03} show substantially different ground-state
properties than approximate solutions, obtained using methods such as
Hartree-Fock (HF) \cite{Hub63,Hir85}, GWF or Gutzwiller Approximation (GA)
\cite{Acq82,Yok87,Lid92,Lid93,Koc99}.
For this reason, computationally-expensive numerical techniques, such as
Quantum Monte Carlo (QMC) \cite{Bec17},
or a~more recent tensor-network
method \cite{Cza16,Sch21,ITe22}, are usually employed for the Hubbard model
in two dimensions, for which exact solution is missing. 

A~notable exception, however, is a~honeycomb lattice, for which relatively
simple techniques, including GWF \cite{Mar97} or CPA \cite{Le13,Row14},
provide reasonable approximations for the critical Hubbard interaction,
differing from the QMC value, $U_c=3.86(1)\,t_0$ \cite{Sor12}
(with $t_0$ being the nearest-neighbor hopping integral and the number in
parenthesis denoting uncertainty for the last digit), by less than $10\%$. 
For a~comparison, the HF method gives $U_c^{\rm (HF)}=2.23\,t_0$ \cite{Sor92}
for the same lattice. 

Since the advent of graphene \cite{Nov05,Zha05} a~half-filled, fermionic
honeycomb-lattice systems have attracted renewed attention, as they
emulate several field-theoretical phenomena in condensed matter
\cite{Kat20}. The effective Hubbard model for monolayer graphene with
on-site interaction $U_{\rm eff}\approx{}1.6\,t_0$ was proposed \cite{Sch13},
suggesting that large isotropic strain may drive this system from semimetallic
towards Mott-insulating phase \cite{Tan15,Zha21} in analogy with high pressure
changing properties of various bulk materials 
\cite{Pas94,Gon05,Dro15,Som19,Cel18}.
Effects of electron correlations are usually more pronounced in graphene
nanosystems, where quantum fluctuations are reduced and magnetic moments
may form near free edges \cite{Fel10,Pot12,Bri22} (although defining
metallic and insulating states for a~nanosystem is more cumbersome than
for a~bulk system \cite{Ryc01,Spa01}). 
We further notice that artificial graphene-like systems allow one to tune
the interaction in a~wider range than actual graphene 
\cite{Sin11,Pol13,Gar20,Tra21}.
Yet another possibility to study electron correlations has open with
the fabrication of twisted bilayer graphene \cite{Cao18,Fid18a}. 

A~separate issue concerns the bandgap opening due to spatial rearrangement
of atoms in strained graphene (a~so-called two-dimensional Peierls
instability), which may turn the system into insulator before the Mott
transition occurs \cite{Lee11,Lee12,Sor18,Eom20,Bao21}.
To the contrary, weak electron-phonon interaction of the Holstein type,
which may appear in graphene on some substrates, is predicted to favor
the semimetallic phase \cite{Cos21}. 

\begin{figure}[!t]
  \includegraphics[width=0.8\linewidth]{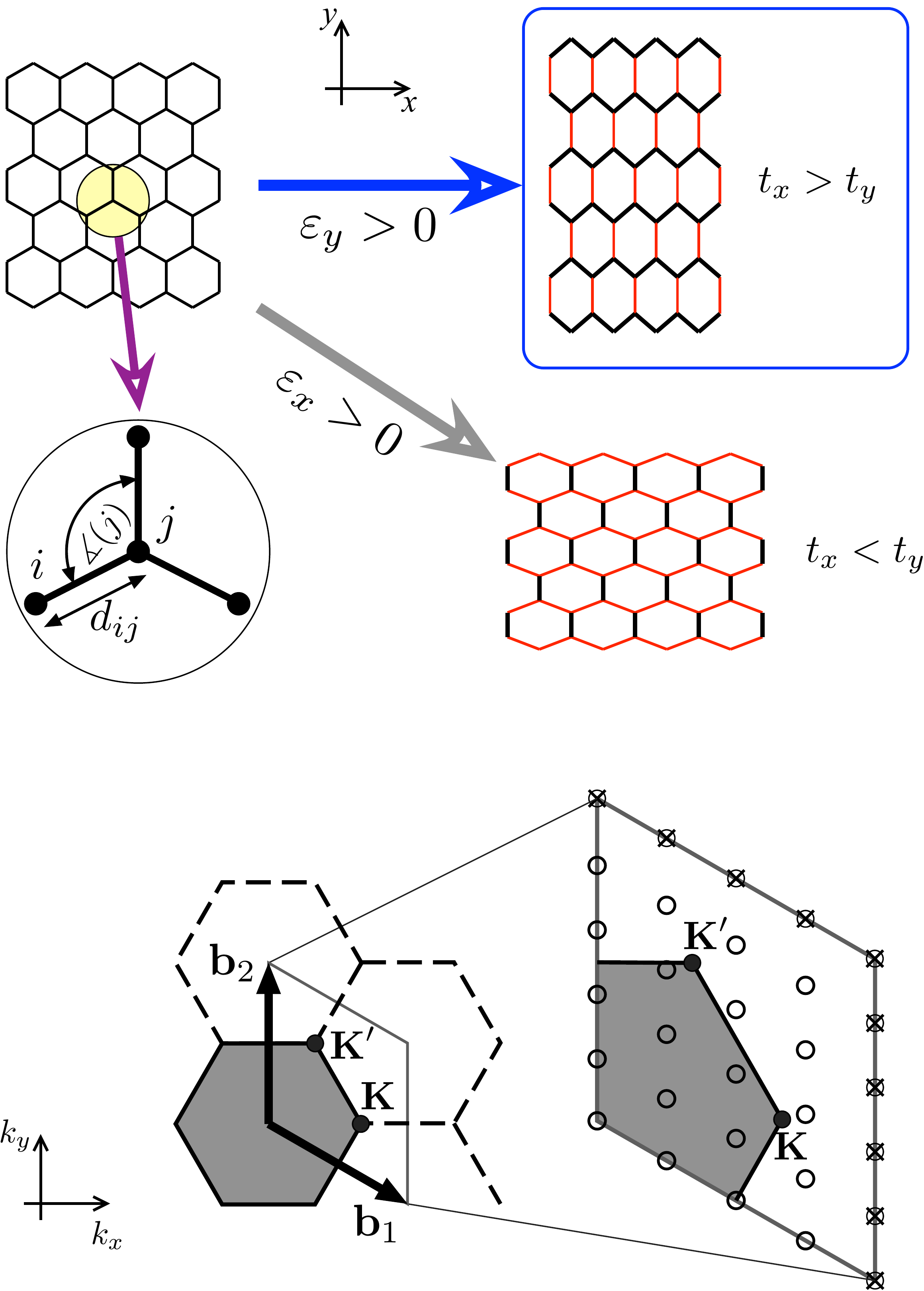}
\caption{ \label{setupxy}
  Top: Honeycomb lattice subjected to uniaxial strain in selected
  direction (see the coordinate system).
  Zoom-in visualizes the distance (bond length) $d_{ij}$ between atoms
  $i$ and $j$ and in-plane angle with the vertex at site $j$
  ($\measuredangle(j)$).
  Bottom: Hexagonal first Brillouin zone (FBZ) of the reciprocal lattice,
  with (dimensionless) basis vectors
  ${\bf b}_1=\left(2\pi/\sqrt{3}\right)\,(\sqrt{3},-1)$ and
  ${\bf b_2}=\left(2\pi/\sqrt{3}\right)\,(0,2)$, and 
  the symmetry points, ${\bf K}=(4\pi/3,0)$ and
  ${\bf K'}=(2\pi/3,2\pi/\sqrt{3})$ coinciding with Dirac points
  in the absence of strain. 
  Magnified area shows discretized FBZ for a~finite system of $N=2N_xN_y$
  atoms with periodic boundary conditions [see Eq.\ (\ref{kxynxy})].
  (The values of $N_x=4$ and $N_y=5$ are used for illustration only.)  
}
\end{figure}

In this paper, the discussion is limited to a~honeycomb lattice strained
along main crystallographic axes (see Fig.\ \ref{setupxy}) supposing that
the bipartite structure of the lattice is preserved under strain.
In turn, there are two different values
of the nearest-neighbor hopping integral in a~single-particle Hamiltonian,
$t_x$ and $t_y$, corresponding to electron hopping along the zigzag
direction ($t_x$) or along the armchair direction ($t_y$).
To obtain a~direct mapping between the strain applied and the hopping
integrals, a~version of the Su-Schrieffer-Heeger (SSH) model
\cite{Dre98} is developed, with microscopic parameters adjusted to match
elastic properties of graphene \cite{Tsa10}. 
Once fixed strain is applied in a~selected direction, the lattice is
allowed to relax along the perpendicular direction to reach a~conditional
energy minimum (a~zero perpendicular stress case).
We further focus our attention on the strain applied along armchair
direction, for which the system evolves towards a~collection of weakly-coupled
one dimensional chains
\cite{Hur01,Spa07,Len16} allowing one to expect that, once the effective
Hubbard model is considered, the Mott transition may appear for smaller value
of $U_{\rm eff}$ than for isotropic strain. 

The remaining part of the paper is organized as follows. 
In Sec.\ \ref{modmets}, we briefly present approximate approaches
to the effective Hubbard Hamiltonian (including HF, GWF, and GA).
Also in Sec.\ \ref{modmets}, we show some original data, illustrating
how these approaches work for anisotropic honeycomb lattice. 
In Sec.\ \ref{resdis}, we discuss our numerical results concerning
the phase diagram of the effective Hubbard model with
arbitrary parameters ($t_x\geqslant{}t_y$ and $U_{\rm eff}$), 
the evolution of the model parameters in graphene subjected to
uniaxial strain, and approximate formula relating the reduction of $U_c$
to strain-induced anisotropy of the Fermi velocity.
The effects of electron correlations on selected measurable quantities
are also presented in Sec.\ \ref{resdis}. 
The concluding remarks are given in Sec.\ \ref{conclu}. 

Next to the main text, in Appendix~\ref{appcpa}, the Coherent Potential
Approximation (CPA) is briefly described. 
In Appendix~\ref{appssh}, we present
the auxiliary SSH model, proposed to relate the physical strain
onto the microscopic parameters of the effective Hubbard model.

\begin{figure}[!t]
  \includegraphics[width=0.9\linewidth]{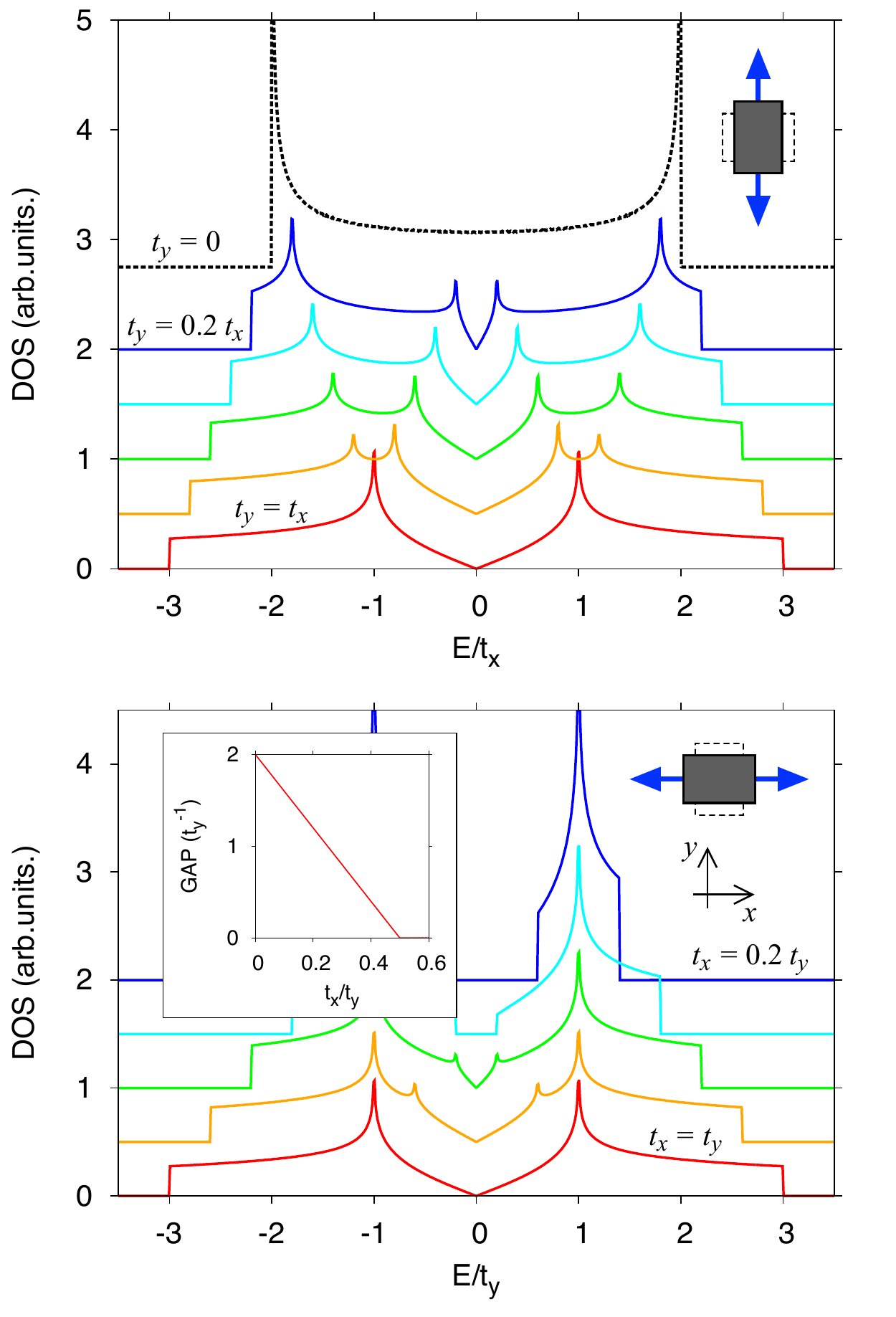}
\caption{ \label{dosplot}
  Density of states for the Hamiltonian (\ref{hamtxyu}) with $U=0$
  displayed as a~function of energy.
  Top: strain applied in the armchair direction ($t_y\leqslant{}t_x$),
  bottom: strain applied in the zigzag direction ($t_y\geqslant{}t_x$).
  The ratio $t_</t_>$ [with $t_<=\mbox{min}\,(t_x,t_y)$ and
  $t_>=\mbox{max}\,(t_x,t_y)$]
  is varied between the lines with the the steps of $0.2$.
  A~vertical offset is applied to each dataset except from the isotropic
  case ($t_x=t_y$).
  Inset shows the band gap, appearing for $t_x<0.5\,t_y$ due to the Peierls
  transition. 
}
\end{figure}

\section{Model and methods}
\label{modmets}

\subsection{The anisotropic Hubbard model} 
Our analysis of electron correlations on anisotropic honeycomb lattice
starts from the Hamiltonian
\begin{equation}
  \label{hamtxyu}
  H = \sum_{\langle{}ij\rangle,s}t_{ij}
  \left(
    c_{i,s}^\dagger{}c_{j,s}+\mbox{H.c.}
  \right) + U\sum_j{} n_{j\uparrow}n_{j\downarrow}, 
\end{equation}
with the first sum running over pairs of nearest-neighbors
$\langle{}ij\rangle$ and spin up/down orientations ($s=\uparrow,\downarrow$), 
and the hopping-matrix elements are given by
\begin{equation}
  \label{tijtxy}
  t_{ij} = 
  \begin{cases}
    -t_x  & \mbox{if $i,j$ belongs to same zigzag line}, \\ 
    -t_y & \mbox{otherwise}. 
  \end{cases}
\end{equation}
(Without loss of generality, we suppose the coordinate system is oriented
as depicted in Fig.\ \ref{setupxy}.)
Remaining symbols in Eq.\ (\ref{hamtxyu}) are a~creation (annihilation)
operator for electron with spin $s$ on the lattice site $i$, $c_{i,s}^\dagger$
($c_{i,s}$), $n_{is}=c_{i,s}^\dagger{}c_{i,s}$, and the on-site Hubbard repulsion $U$.
We further limit our considerations to the ground state and suppose the
half-filling, i.e., one electron per lattice site,
$\overline{n}=\langle{}n_{i\uparrow}+n_{i\downarrow}\rangle=1$. 

In principle, ground-state properties of the model defined by Eqs.\
(\ref{hamtxyu}) and (\ref{tijtxy}) can be discussed as functions of two
dimensionless parameters, e.g., $t_y/t_x$ and $U/t_x$. 
The relation between parameters $t_x$ and $t_y$ and strain applied to graphene
is discussed later in this section.
But first, we briefly present approximate approaches capable to distinguish
whether ground state of the Hamiltonian (\ref{hamtxyu}) is semimetallic
or insulating.

\subsection{Hartree-Fock approximation}
Although a~honeycomb lattice is bipartite and the antiferromagnetic order
is possible, its peculiar band structure suppresses antiferromagnetism
at small $U$ \cite{Mar97}.
Since a~single particle density of states (i.e, density of states at $U=0$)
is linear for low energies, see Fig.\ \ref{dosplot}, there is no Fermi
surface that could produce magnetic instability also for small $U>0$.

Within the Hartree-Fock approximation, interaction part in the Hamiltonian
(\ref{hamtxyu}) is replaced by
\begin{equation}
  U\hat{D} \stackrel{\rm HF}{=} U\sum_i\left(
    \langle{}n_{i\uparrow}\rangle{}n_{i\downarrow} +
    n_{i\uparrow}\langle{}n_{i\downarrow}\rangle -
    \langle{}n_{i\uparrow}\rangle{}\langle{}n_{i\downarrow}\rangle
  \right), 
\end{equation}
where we have introduced the operator $\hat{D}=\sum_jn_{j\uparrow}n_{j\downarrow}$
measuring the number of double occupancies.
We further impose the antiferromagnetic order,
\begin{equation}
\label{nis0af}
   \langle{}n_{i\uparrow}\rangle = \frac{\overline{n}+\lambda_i{}m}{2},
   \ \ \ \ \ \ 
   \langle{}n_{i\downarrow}\rangle = \frac{\overline{n}-\lambda_i{}m}{2}, 
\end{equation}
where $\lambda_i=1$ if $i$ belongs to one sublattice ($A$), or
$\lambda_i=-1$ if $i$ belongs to the other sublattice ($B$), and
$m$ is the magnetization ($|m|\leqslant\overline{n}$), and the half filling
($\overline{n}=1$).
The above yields the HF ground-state energy per site
\begin{align}
  \frac{E_G^{{\rm (HF)}}}{N} &=
  -\frac{2}{N}\sum_{\bf k}\sqrt{E_{\bf k}^2+\left(\frac{Um}{2}\right)^2}
  + \frac{U(1\!+\!m^2)}{4}, 
  \label{eghf}
\end{align}
where the factor $2$ accounts for $s=\uparrow,\downarrow$, and
the summation runs over quasimomenta ${\bf k}\equiv{}(k_x,k_y)$
in the first Brillouin zone, namely
\begin{align}
  k_x &= \frac{2\pi{}}{N_x}n_x,
  \ \ \ \ \ \ 
  k_y = \frac{4\pi{}}{\sqrt{3}}\left(\frac{n_y}{N_y}-\frac{n_x}{2N_x}\right),
  \label{kxynxy}
  \\
  n_x &= 0,1,\dots,N_x\!-\!1,  \ \ \ \ n_y = 0,1,\dots,N_y\!-\!1,
  \nonumber
\end{align}
with $N_{x,y}$ being the number of unit cells in $x,y$ direction, $N=2N_xN_y$
(the periodic boundary conditions are imposed). 
For sufficiently large number of points in the momentum space, say
$N_x,N_y\gtrsim{}10^3$, one can usually works with a~square inverse lattice
(omitting the term $\propto{}n_x$ in the expression for $k_y$) \cite{kxyfoo};
nevertheless, the discretization of $(k_x,k_y)$ as given in Eq.\ (\ref{kxynxy})
becomes crucial when discussing the finite-size effects for small $N$. 
The single-particle energies for anisotropic honeycomb lattice are given by
\begin{equation}
  \label{epsilonk}
  E_{\bf k} = t_x\sqrt{a_{\bf k}^2+b_{\bf k}^2}, 
\end{equation}
with
\begin{align}
  a_{\bf k} =&
    -\cos\left(\frac{k_x}{2}+\frac{\sqrt{3}k_y}{2}\right)
    -\cos\left(\frac{k_x}{2}-\frac{\sqrt{3}k_y}{2}\right) - \frac{t_y}{t_x},
   \nonumber \\
  b_{\bf k} =& 
    \sin\left(\frac{k_x}{2}+\frac{\sqrt{3}k_y}{2}\right)
    -\sin\left(\frac{k_x}{2}-\frac{\sqrt{3}k_y}{2}\right).
\end{align}

Next, the density of states is defined as 
\begin{equation}
  \label{rhodef}
  \rho(E) = 2{N}^{-1}\sum_{\bf k}\left[
    \delta\left(E-E_{\bf k}\right)+
    \delta\left(E+E_{\bf k}\right)
  \right], 
\end{equation}
with two parts corresponding to the conduction ($E>0$) and
valence ($E<0$) band, and satisfying the normalization conditions: 
$\int_{-\infty}^0{}dE\,\rho(E)=\int_0^{\infty}dE\,\rho(E)=1$.
If $m\neq{}0$, the minimization of $E_G^{\rm (HF)}$ given by
Eq.\ (\ref{eghf}) brought us to 
\begin{equation}
\label{gapeq1}
  1 = \int_{E<0}dE\,\rho(E)\frac{U/2}{\sqrt{E^2+(Um/2)^2}}. 
\end{equation}
In case the solution of Eq.\ (\ref{gapeq1}) does not exist, the  minimum
of $E_G^{\rm (HF)}$ corresponds to $m=0$. 

Unlike for square lattice, for which one gets $m\neq{}0$ for any $U>0$
\cite{Hir85}, on a~honeycomb lattice the minimization gives $m=0$ for
$U\leqslant{}U_c^{\rm (HF)}$ and $m\neq{}0$ for $U>U_c^{\rm (HF)}$ \cite{Mar97}. 
This can be easily understood for the case of unstrained (or uniformly
strained) lattice, for which $t_x=t_y=t_0$ and the density of states can
be approximated by
\begin{equation}
\label{toydos}
  \rho(E)\approx{}\rho_\Lambda(E) =
  \begin{cases}
    \frac{2}{\Lambda{}^2}|E| & \text{for }\ |E|\leqslant{}\Lambda,  \\
    0 & \text{for }\ |E|>\Lambda, \\
  \end{cases}
\end{equation}
with a~cut-off energy of 
$\Lambda=\left(\sqrt{3}\pi\right)^{1/2}\,t_0\simeq{}2.33268\,t_0$.
The above is equivalent, for $|E|\leqslant{}\Lambda$, to
$\rho_\Lambda(E)=2{\cal A}|E|/\left[N\pi(\hbar{}v_F)^2\right]$,
with ${\cal A}$ being the
system area, $v_F=\frac{1}{2}\sqrt{3}\,at_0/\hbar$ the Fermi velocity, and
$a$ the lattice parameter \cite{Cas09}. 
It is straightforward to show that $m\neq{}0$ appears above
$U_c^{(\Lambda)}=\Lambda$, being not far from the value reported in Ref.\
\cite{Sor92}. 

The values of $U_c^{\rm (HF)}$ following from numerical minimization of
$E_G^{\rm (HF)}$ given by Eq.\ (\ref{eghf}) for the actual density of states
are presented in Sec.\ \ref{resdis}.

\begin{figure}[!t]
  \includegraphics[width=\linewidth]{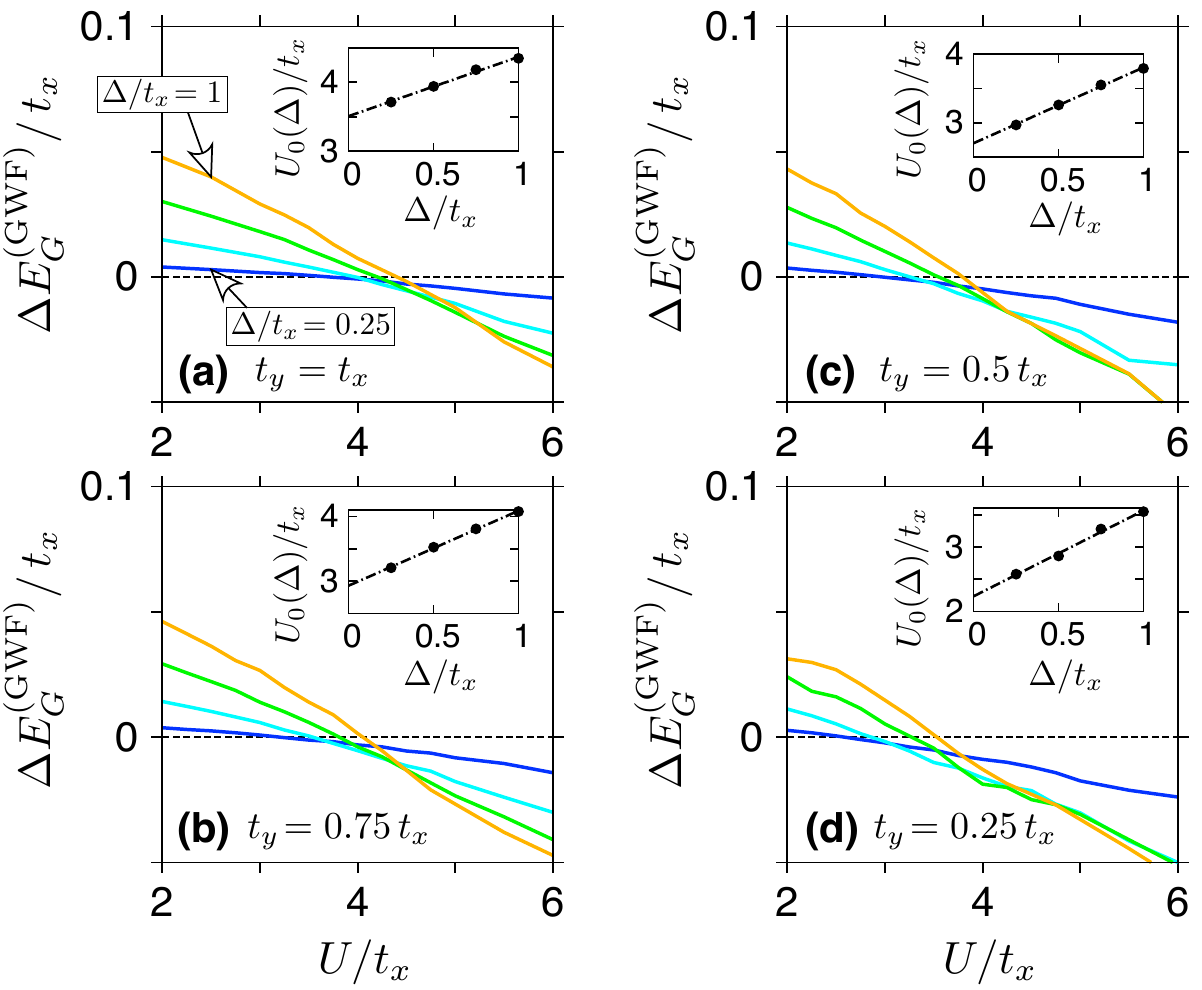}
\caption{ \label{denvsu}
  (a)--(d) Main: Energy difference between the antiferromagnetic Gutzwiller
  variational energy $E_G^{\rm (GWF)}(m)$ [see Eq.\ (\ref{eggwf})] and
  the paramagnetic solution $E_G^{\rm (GWF)}(0)$ obtained from VMC simulations
  as a~function the on-site Hubbard repulsion ($U$). The parameter $\eta$ is
  optimized for a~fixed $m=\Delta/U$ (or $m=0$);
  $\Delta$ is varied between the lines
  from $\Delta/t_x=0.25$ to $\Delta/t_x=1$, with the steps of $0.25$.
  The hopping anisotropy $t_y/t_x$ is varied between the panels.
  Inset shows the value of $U=U_0(\Delta)$ at which $\Delta{}E_G^{\rm (GWF)}$
  changes sign for a~given $\Delta$. 
  The extrapolation to $\Delta\rightarrow{}0$ yields the critical values
  of $U_{\rm c}^{\rm (GWF)}$ given in Table~\ref{uctable}. 
  (Statistical errorbars are to small to be shown on the plots.) 
}
\end{figure}

\subsection{Gutzwiller wavefunction}
Generalized Gutzwiller wavefunction, allowing antiferromagnetic order,
was applied in Ref.\ \cite{Mar97} to find out that correlated, but
paramagnetic solution remains stable up to the region of the Mott
semimetal-insulator transition.
Although several features of the solution are altered when employing
more advanced techniques \cite{Sor12}, the values of $U_c$ following
from GWF are surprisingly close to those obtained within large-scale
computer simulations for a~honeycomb lattice. 
Investigating the variational wavefunction 
\begin{equation}
\label{psigwf}
\left|\Psi_{\rm GWF}\right\rangle=e^{-\eta\hat{D}}\left|\psi_0(m)\right\rangle, 
\end{equation}
where $\left|\psi_0(m)\right\rangle$ denotes a~Slater determinant
corresponding to a~given magnetization in Eq.\ (\ref{eghf}) and $\eta$
is another variational parameter (quantifying the role of electron
correlations), one needs to minimize the ground-state energy
\begin{equation}
\label{eggwf}
  E_G^{\rm (GWF)}= \frac{\left\langle{}
  \psi_0(m)\right|e^{-\eta\hat{D}} H e^{-\eta\hat{D}}\left|\psi_0(m)
  \right\rangle}{\left\langle
  \psi_0(m)\right|e^{-2\eta\hat{D}}\left|\psi_0(m)
  \right\rangle}, 
\end{equation}
with respect to $\eta$ and $m$.
In many cases, the system may prefer
to reduce $m$ (even to $m=0$) and increase $\eta$, allowing to expect that,
in general, $U_c^{\rm (GWF)}\geqslant{}U_c^{\rm (HF)}$. 

Several approximated techniques for calculating the averages in
Eq.\ (\ref{eggwf}) were developed \cite{Yok87,Lid92,Lid93,Mar97,Koc99}.
Here, we apply Variational Monte Carlo (VMC), described in details in Ref.\
\cite{Koc99}. To determine the value of $U_c^{\rm (GWF)}$, we have directly
followed the procedure proposed by Martelo {\it et al.\/} \cite{Mar97}.
For a~fixed value of the gap ($\Delta\equiv{}Um$), the energy difference
$\Delta{}E_G^{\rm (GWF)}=E_G^{\rm (GWF)}(m)-E_G^{\rm (GWF)}(0)$, where
the parameter $\eta$ is optimized independently for $m=0$ and $m\neq{}0$,
changes sign at some $U=U_0(\Delta)$. Numerical extrapolation of
$U_0(\Delta)$ with $\Delta\rightarrow{}0$ allows one to determine the
critical value of $U_c^{\rm (GWF)}$.
Selected examples, for $t_y\leqslant{}t_x$ (i.e., strain applied in
the armchair direction) and the system size of $N=200$ sites ($N_x=N_y=10$), 
are presented in Fig.\ \ref{denvsu}. 
For more details of the simulation, see Ref.\ \cite{vmcfoo}.

\begin{table}[!b]
\caption{ \label{uctable}
  Critical values of the Hubbard repulsion $U_c^{\rm (GWF)}$ obtained from
  VMC simulations (with standard deviations for the last digit specified
  in parentheses) compared with the upper ($U_c^{\rm (GA)}$) and the upper
  ($U_c^{\rm (NGA)}$) bound following from the Gutzwiller Approximation (GA)
  and the N\'{e}el-state Gutzwiller Approximation (NGA).
  The results obtained from the Statistically-consistent Gutzwiller
  Approximation (SGA) are also given. 
  The system size is defined by $N_x=N_y=10$ for VMC simulations;
  remaining results correspond to the limit of $N_x=N_y\rightarrow{}\infty$. 
}

\vspace{0.5em}
\begin{tabular}{c|cccc}
\hline\hline
$\ \ t_y/t_x\ \ $ & $\ U_c^{\rm (GWF)}\!/t_x\ $ & $\ U_c^{\rm (GA)}\!/t_x\ $
  &  $\ U_c^{\rm (SGA)}\!/t_x\ $ & $\ U_c^{\rm (NGA)}\!/t_x\ $  \\
\hline
1.00 & 3.48(1) & 2.804 & 3.122 & 5.281 \\
0.75 & 2.91(1) & 2.550 & 2.833 & 4.871 \\
0.50 & 2.69(3) & 2.241 & 2.468 & 4.508 \\
0.25 & 2.24(1) & 1.830 & 1.983 & 4.199 \\
\hline\hline
\end{tabular}
\end{table}

\subsection{Gutzwiller Approximation and its variants}
To efficiently study the effects of electron correlations present
in $|\Psi_{\rm GWF}\rangle$, see Eq.\ (\ref{psigwf}), one can also adopt the
Gutzwiller Approximation (GA) and find out how the number of double
occupancies is reduced comparing to the HF solution $|\psi_0(m)\rangle$.
Within GA, which is exact in the infinite dimension limit, the correlation
functions
$
\langle{}c_{is}^{\dagger}c_{js}\rangle_{\rm GWF}=\left\langle{}
\Psi_{\rm GWF}\right|c_{is}^{\dagger}c_{js}\left|\Psi_{\rm GWF}\right\rangle/
\left\langle\Psi_{\rm GWF}\right|\left.\Psi_{\rm GWF}\right\rangle
$
are approximated by
\begin{equation}
  \langle{}c_{is}^{\dagger}c_{js}\rangle_{\rm GWF}
  \stackrel{\rm GA}{=}
  q(\{\rho_{ll's'}^{(0)}\};\{d_l\})\,\langle{}c_{is}^{\dagger}c_{js}\rangle_0
\end{equation}
where the band-narrowing factor $q(\{\rho_{ll's'}^{(0)}\};\{d_l\})$
depends only on the single-particle density matrix elements
$\rho_{ll's'}^{(0)}=\langle{}c_{ls'}^{\dagger}c_{l's'}\rangle_0$ with $l,l'=i,j$ and  
$s'=\uparrow,\downarrow$ (here, $\langle{}\dots\rangle_0$ is the expectation
value over the uncorrelated state; i.e., a~single Slater determinant such
as $|\psi_0(m)\rangle$),
and $d_l=\langle{}n_{l\uparrow}n_{l\downarrow}\rangle_{\rm GWF}$  ($l=i,j$)
being the average double occupancies.
The $\{d_i\}$ variables are further regarded as variational parameters
to be determined by minimizing the Gutzwiller energy functional,
\begin{equation}
  E_G^{\rm (GA)}=2\sum_{\langle{}ij\rangle,s}q(\{\rho_{ll's'}^{(0)}\};\{d_l\})
  \,t_{ij}\rho_{ijs}^{(0)} + U\sum_jd_j. 
\end{equation}

Several forms of the band-narrowing factor
$q(\{\rho_{ll's}^{(0)}\};\{d_l\})$,
being equivalent in the infinite dimension
limit but producing slightly different results when applied to the system
of a~finite dimensionality, are used among the literature
\cite{Lid92,Tak75,Vol84,Jed10,Lan12,Wys14,Che17,Fid18b}. 
For the diagonal elements $\rho_{iis}^{(0)}=\langle{}n_{is}\rangle_0$
parametrized as in Eq.\ (\ref{nis0af}) with $\overline{n}=1$, one can
impose $d_i\equiv{}d$ for all sites and rewrite the expression given
in Ref.\ \cite{Che17} as
\begin{align}
  q(\{\rho_{ll's}^{(0)}\};\{d_l\}) &\equiv{} q(m,d) = \nonumber \\
   &
   \frac{4d}{1\!-\!m^2}\left[
    1\!-\!2d+\sqrt{(1\!-\!2d)^2-m^2} 
  \right].
  \label{qumd}
\end{align}
The variable $d$ is bounded as $0\leqslant{}d\leqslant{}\frac{1}{4}
(1\!-\!m^2)$, with the upper limit corresponding to the average double
occupancy in the uncorrelated state $|\psi_0(m)\rangle$). 
The kinetic energy term can be estimated by referring to the Hatree-Fock
energy $E_G^{\rm (HF)}$, see Eq.\ (\ref{eghf}), as
$
2\sum_{\langle{}ij\rangle{},s}t_{ij}\rho_{ij}^{(0)} =
E_G^{\rm (HF)}-\frac{N}{4}U(1-m^2)
$
even for $m$ being away from the minimum of $E_G^{\rm (HF)}$. 
This brought us to
\begin{align}
  \frac{E_G^{\rm (GA)}}{N} &= q(m,d)\times \nonumber \\
  &
  \left[
    -\frac{2}{N}\sum_{\bf k}\sqrt{E_{\bf k}^2+\left(\frac{Um}{2}\right)^2}
    + \frac{Um^2}{2}
  \right] + Ud.
  \label{eggamd}
\end{align}

Numerical minimization of $E_G^{\rm (GA)}$, with respect to $(m,d)$,
truncates the optimization of both the density matrix
$\{\rho_{i,j}^{(0)}\}$ and the parameters $\{d_i\}$.
For the linear density of states $\rho_{\Lambda}(E)$, see Eq.\ (\ref{toydos}),
one can easily find closed-from expression for $E_G^{\rm (GA)}$; 
the minima corresponding to $m\neq{}0$ appear for
$U>U_c^{(\Lambda,{\rm GA})}=1.270\,\Lambda = 2.963\,t_0$, with the critical
value lying between HF \cite{Sor92} and QMC \cite{Sor12} results for
isotropic honeycomb lattice. 

A~slightly more accurate (but also more computationally expensive) approach
can be constituted by parametrizing the uncorrelated state
$|\psi_0\rangle$ not only via the magnetization $m$, as in the above, but
via {\em all} independent parameters of the density matrix
$\{\rho_{i,j}^{(0)}\}$. In particular, the auxiliary single-particle Hamiltonian
determining $\{\rho_{i,j}^{(0)}\}$ contains the renormalized hopping integrals
($\tilde{t}_x$ and $\tilde{t}_y$) which may differ from $t_x$ and $t_y$ in
the multiparticle Hamiltonian (\ref{hamtxyu}). 
The resulting method, called the
{\em Statistically-consistent Gutzwiller Approximation} (SGA), is presented
in details in Ref.\ \cite{Jed10}. 

Both (S)GA and GWF methods can be regarded as improvements to mean-field
(HF) solution, including some classes of quantum fluctuations.
Since not all fluctuations are included, the AF order is artificially
favored when searching for the energy minimum, and therefore these
methods usually underestimate the value of $U_c$.
In order to bound $U_c$ from the top, we employ the scheme proposed
by Martelo {\em et al.\/} \cite{Mar97}, in which two solutions are
compared: The~paramagnetic GA solution, corresponding $m=0$ in
Eq.\ (\ref{eggamd}), with a~complementary variational
wavefunction,
\begin{equation}
\label{psibe}
  |\Psi_B\rangle = e^{-\kappa{}\hat{T}}|\Psi_{U\rightarrow\infty}\rangle, 
\end{equation}
where $\kappa$ is a~variational parameter, $\hat{T}=\sum_{\langle{}ij\rangle{},s}
t_{ij}(c_{is}^{\dagger}c_{js}+\mbox{H.c.})$ is the kinetic-energy part of the
Hamiltonian (\ref{hamtxyu}), and $|\Psi_{U\rightarrow\infty}\rangle$ is the
ground state for $U\rightarrow\infty$.
The critical value $U_c$ is than estimated by finding a~crossing point
of $E_G^{\rm (GA)}$, Eq.\ (\ref{eggamd}), with a~fixed $m=0$ and optimized $d$,
and the variational energy $E_B$ corresponding $|\Psi_B\rangle$,
Eq.\ (\ref{psibe}), with optimized $\kappa$. 

For $m=0$, the factor $q(m,d)$, Eq.\ (\ref{qumd}), reduces to a~quadratic
function of $d$ and the functional $E_G^{\rm (GA)}$, Eq.\ (\ref{eggamd}),
reaches the minimum at
$d=\frac{1}{4}\mbox{max}\left[0,1-U/(8|\epsilon_0|)\right]$, leading to
a~form originally derived by Gutzwiller \cite{Gut64,Gut65}
\begin{equation}
  \label{eggam0}
  E_G^{\rm (GA)}(m\!=\!0) = \begin{cases}
  {\displaystyle\epsilon_0+\frac{U}{4}-\frac{U^2}{64|\epsilon_0|}}
  & \text{for } U\leqslant{8|\epsilon_0|}, \\
  0 & \text{otherwise}. 
  \end{cases}
\end{equation}
The symbol $\epsilon_0$ is the kinetic energy per site for $U=0$,
namely $\epsilon_0=\int_{E<0}dE{}\,\rho(E)E$ [for the definition of $\rho(E)$,
see Eq.\ (\ref{rhodef})], taking the numerical value of
$\epsilon_0/t_x=-1.57460$ for $t_y/t_x=1$, $\epsilon_0/t_x=-1.45540$
for $t_y/t_x=0.75$, $\epsilon_0/t_x=-1.36218$ for $t_y/t_x=0.5$, or
$\epsilon_0/t_x=-1.29891$ for $t_y/t_x=0.25$.

\begin{figure}[!t]
  \includegraphics[width=\linewidth]{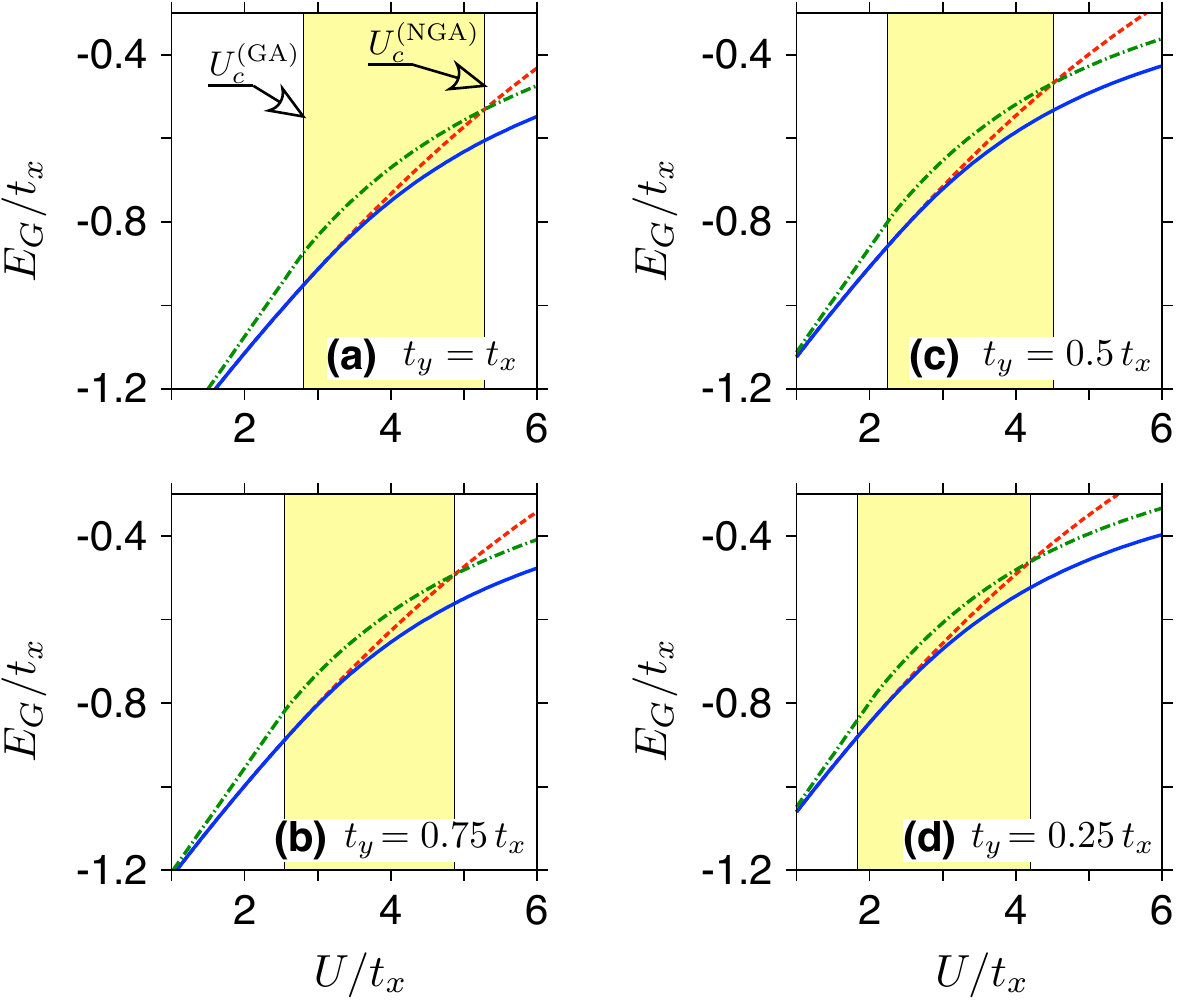}
\caption{ \label{gangafig}
  (a)--(d). The lower and the upper bounds to the critical Hubbard repulsion
  $U_c$ for anisotropic honeycomb lattice estimated by comparing different
  versions of the Gutzwiller Approximation described in the text. 
  The lower bound ($U_c^{\rm (GA)}$) coincides with the splitting of the
  Gutzwiller energy for paramagnetic state, $E_G^{\rm (GA)}(m\!=\!0)$,
  given by Eq.\ (\ref{eggam0}) [red dashed line] and the variational energy
  $E_G^{\rm (GA)}$, see Eq.\ (\ref{eggamd}), with the parameters $(m,d)$
  optimized numerically [blue solid line], both displayed as functions of $U$.
  The value of $U_c^{\rm (GA)}$ is obtained via the extrapolation
  with $m\rightarrow{}0$, similarly as for the VMC results
  in Fig.\ \ref{denvsu}.
  The intersection of $E_G^{\rm (GA)}(m\!=\!0)$ with $E_G^{\rm NGA}$, see
  Eq.\ (\ref{egganga}) [green dashed-dotted line] yields the upper bound
  ($U_c^{\rm (NGA)}$).
  The value of $t_y/t_x$ ratio is varied between the panels.
  [For the numerical values of $U_c^{\rm (GA)}$ and $U_c^{\rm (NGA)}$, see
  Table~\ref{uctable}.]
}
\end{figure}

In the limit of infinite dimensions, the variational energy $E_B$
associated with the state $|\Psi_B\rangle$ can be evaluated exactly,
since the ground state $|\Psi_{U\rightarrow{}\infty}\rangle$ is the N\'{e}el
antiferromagnet \cite{Ken88}.
The variational energy reads
\begin{equation}
  \label{egganga}
  \frac{E_B}{N} = \epsilon_{\rm kin}+\frac{U(1\!-\!m^2)}{4}\equiv{}
  \frac{E_G^{\rm (NGA)}}{N}
\end{equation}
where
\begin{align}
   \epsilon_{\rm kin} &= \int_{E<0}dE\,\rho(E)E
   \left[-\tanh(2\kappa{}E)\right],
   \label{ekinga}
   \\
   m &= \int_{E<0}dE\,\frac{\rho(E)}{\cosh(2\kappa{}E)},
   \label{mmnga}
\end{align}
are the kinetic energy per site and the sublattice magnetization
(respectively). 
The so-called N\'{e}el-Gutzwiller Approximation (NGA) is constituted by
substituting the density of states given by Eq.\ (\ref{rhodef}) into
Eqs.\ (\ref{ekinga}), (\ref{mmnga}), and the subsequent minimization
of $E_B\equiv{}E_G^{\rm (NGA)}$ with respect to $\kappa$. Selected numerical
results, for $t_y\leqslant{}t_x$, are presented in Fig.\ \ref{gangafig}.

It is worth mentioning that (S)GA can be systematically improved, approaching
the GWF solution, by including consecutive corrections following from the
relevant diagrammatic expansion \cite{Lid92,Wys14,Fid18b}
(we further notice that the detailed scheme for a~honeycomb lattice
is missing so far).
Similar approach for $|\Psi_B\rangle$ is difficult due to necessity
of determining the ground state of the Heisenberg model
($|\Psi_{U\rightarrow{}\infty}\rangle$) as a~first.

Selected numerical values of $U_c$, following from the methods described
in this Section, are compared in Table~\ref{uctable}. 

A~substantially different approach, the {\em Coherent Potential Approximation}
(CPA), in which one considers random scattering of electrons with a~given spin
on motionless electrons with the opposite spin (instead of imposing some spin
order), is described in Appendix~\ref{appcpa}.

\begin{figure}[!t]
  \includegraphics[width=\linewidth]{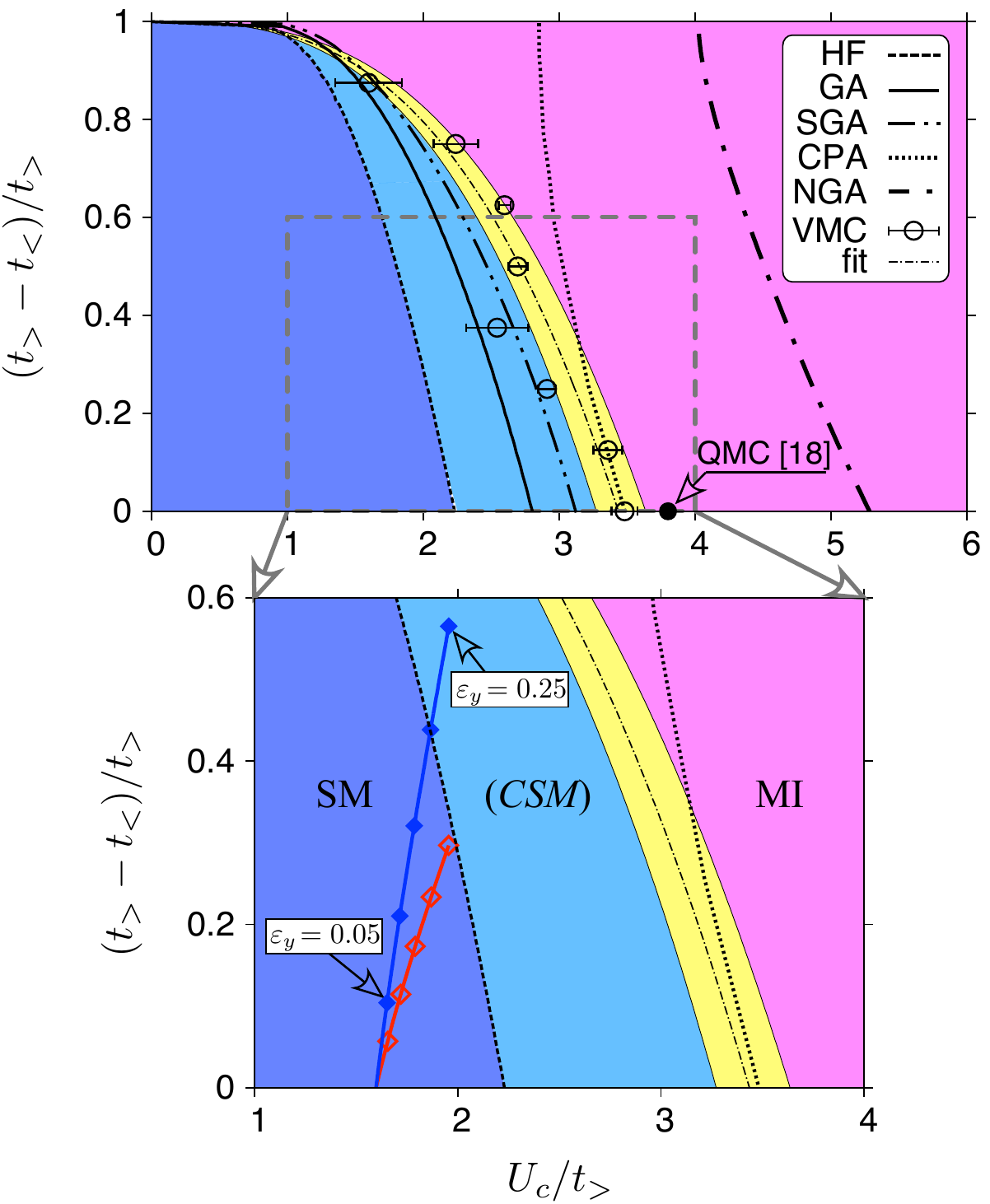}
  \caption{ \label{phadiag2}
    Top: Phase diagram for the Hubbard model on anisotropic honeycomb lattice,
    see Eqs.\ (\ref{hamtxyu}) and (\ref{tijtxy}), with $t_y\leqslant{}t_x$
    corresponding a~gapless single-particle spectrum, see Fig.\ \ref{dosplot}.
    [Here, $t_>=t_x$ and $t_<=t_y$.]
    Lines depict the critical Hubbard repulsion estimated within the
    Hartree-Fock method [short dashed], Gutzwiller Approximation
    [thick solid],
    Statistically-consistent GA [long dashed-double dotted],  
    Coherent Potential Approximation [dotted],
    and the N\'{e}el-state GA [long dashed-dotted].
    Datapoints with errorbars are obtained from VMC simulations for the
    Gutzwiller Wave Function, see Eq.\ (\ref{psigwf}); thin dash-dotted line
    represents a~power-law fit given by Eq.\ (\ref{uctxyfit}) with thin solid
    lines bounding the statistical uncertainty (yellow area).
    Quantum Monte Carlo value for the isotropic case, $U_c/t_0=3.86$
    \cite{Sor12}, is also mark (full circle). 
    Bottom: A~zoom in, with trajectories following from the SSH model for
    strained graphene (see Appendix~\ref{appssh}) for $\beta=2$ (red line/open
    symbols) and $\beta=3$ (blue line/closed symbols).
    Different datapoints for each value of $\beta$ correspond to the applied
    strain $\varepsilon_y$ varied from $\varepsilon_y=0.05$ to $0.25$ with
    the steps of $0.05$. (GA and VMC results are omitted for clarity.)
    Remaining Labels/colored~areas:
    the semimetallic phase (SM) [blue] with the correlated-semimetal range
    ({\em CSM}) [light blue] and the Mott insulator (MI) [magenta]. 
  }
\end{figure}

\section{Results and discussion}
\label{resdis}

\subsection{Phase diagram}
Our central results are presented in Fig.\ \ref{phadiag2}, where we display
the phase diagram for the Hamiltonian (\ref{hamtxyu}) with the strain
applied in armchair direction ($t_y\leqslant{}t_x$). 
A~single-particle spectrum is gapless in such a~case (see also Fig.\
\ref{dosplot}) since the positions of Dirac cones do not merge \cite{Per09};
therefore, metal-insulator (if occurs) must be driven by
electron-electron interaction.
Most of the methods which we have presented in Sec.\ \ref{modmets}, i.e.,
HF, GA, and NGA, share a~common feature that they allows one to take
the limit of $N\rightarrow{}\infty$ numerically, and the results are free of
finite-size (and statistical) errors.
Same applies to SGA (see Ref.\ \cite{Jed10}) and CPA described in
Appendix~\ref{appcpa}. 
The case of GWF is different, since VMC simulations produced considerable
statistical errorbars
(a~triple standard deviation is marked for each datapoint) and may be biased
due to possible systematic errors following from a~limited system size
of $N=200$ ($N_x=N_y=10$). 

Despite the limited accuracy of VMC simulation results, they typically
lie between the GA and CPA values (up to the errorbars), allowing to
regard the last to methods as providing approximate lower (GA) and upper
(CPA) bounds to the value of $U_c$. However, it must be noticed that the
'exact' numerical value of $U_c^{\rm (QMC)}=3.86\,t_0$ of Ref.\ \cite{Sor12}
(available only for the isotropic case, $t_x=t_y=t_0$) significantly exceeds
$U_c^{\rm (CPA)}=3.49\,t_0$, and therefore the CPA results cannot be considered as
upper bound to $U_c$ in a~rigorous manner.
When searching for a~computationally-inexpensive technique providing
a~safe upper bound to $U_c$, one should rather refer to Ne\'{e}l-state
Gutzwiller Approximation (NGA).

The relation between SGA and the above-mentioned methods is more complex,
since more variational parameters defining the single-particle state
$|\psi_0\rangle$ are optimized.
In brief, the SGA ground-state energy lower or equal than the
obtained from GA, leading to $U_c^{\rm (SGA)}\geqslant{}U_c^{\rm (GA)}$.
However, the mutual relation between SGA and VMC results cannot be
determined {\em a~priori}, as the former provides better optimization of
$|\psi_0\rangle$, whereas the latter put more emphasize on accurate
calculation of averages in Eq.\ (\ref{eggwf}).
Looking at the results presented in Fig.\ \ref{phadiag2}, we may conclude
that $U_c^{\rm (GWF)}\gtrsim{}U_c^{\rm (SGA)}$, finding SGA as slightly
less accurate, but promising (due to much lower computational costs)
counterpart to VMC. 

The VMC results concerning $U_c$ can be rationalized within a~power law,
with least-square fitted parameters, as follows
\begin{equation}
\label{uctxyfit}
  \frac{t_y}{t_x} = (0.0269\pm{}0.0014)\times
  \left(\frac{U_c}{t_x}\right)^{2.93\pm{}0.05}. 
\end{equation}
(Here, a~single standard deviation is given for each parameter.)
The line given by Eq.\ (\ref{uctxyfit}) [dashed-dotted], surrounded by
the area [yellow] marking the statistical uncertainty, is further regarded
as a~border between semimetallic (SM) and Mott-insulating (MI) regions in
the phase diagram.
The former is further divided by marking the correlated-semimetal range
(CSM), an appearance of which can be attributed to the fact that
the HF approximation no longer produces a~correct paramagnetic solution
($m=0$).
Such a~computation-oriented notion cannot be regarded as a~thermodynamic
phase {\em per se}; however, prominent effects of electron correlations, i.e.,
the band narrowing and the reduction of double occupancies are gradually
amplified when the interaction is increased.
These effects are further discussed in next subsection, where we describe
the behavior of measurable quantities when passing the CSM range and
approaching the metal-insulator boundary. 

Three of the methods (HF, GA, and GWF) indicate $U_c\rightarrow{}0$ for
$t_y/t_x\rightarrow{}0$, coinciding with the exact solution for the Hubbard
chain \cite{Lie68,Lie03}, giving an insulating phase at arbitrarily small
$U>0$. In contrast, CPA and NGA produces $U_c>0$ in such a~limit, showing
that these are inapplicable in the limit of weakly-coupled chains, despite
producing a~reasonable results in the isotropic case. (In particular,
when comparing to the value of $U_c/t_0=10$ given by DMFT \cite{Tra09}.)

Two striking features of the data shown in Fig.\ \ref{phadiag2} are
that most of the VMC datapoints do not match the GA line within the errorbars,
but --- on the other hand --- the points for $t_y/t_x<0.8$ match the CPA
results surprisingly close.
The above may indicate a~role of finite-size effects in VMC simulations
(notice that both GA and CPA solutions correspond to the
$N\rightarrow\infty$ limit). By manipulating the system sizes used
for HF and GA calculations we found that shrinking to $N_x=N_y=10$ usually
produces $U_c/t_x$ enlarged by $0.1$ (HF) or $0.2$ (GA) comparing to the
large-system limit. Therefore, one could roughly estimate $U_c^{\rm (GWF)}/t_x$
to be reduces by $0.2-0.3$ when enlarging the system for $t_y\approx{}t_x$.
This quantity is comparable but smaller than the deviation from the 'exact'
QMC result of Ref.\ \cite{Sor12}, namely
$U_c^{\rm (QMC)}-U_c^{\rm (GWF)}\approx{}0.4\,t_x$, suggesting that, in search
for more accurate VMC results, one should first include additional
variational parameters (such as Jastrow factors \cite{Cap05,Bib18}), while
the role of system size is rather secondary. 

Also in Fig.\ \ref{phadiag2} (bottom panel) we depict the trajectories
followed by a~sheet of graphene subjected to a~strain in armchair direction
($\varepsilon_y>0$) and allowed to relax in the perpendicular (i.e., zigzag)
direction. The hopping matrix elements in the Hamiltonian (\ref{hamtxyu})
are parametrized according to \cite{Dre98,Gro18,Ryc13}
\begin{equation}
\label{tijdij}
  t_{ij} = -t_0\left(1-\beta\frac{\delta{}d_{ij}}{d_0}\right), 
\end{equation}
where $\beta=2$ (red solid line; open symbols) or $\beta=3$ (blue solid line;
closed symbols) is the dimensionless electron-phonon coupling parameter. 
The bond-length variations ($\delta{}d_{ij}$) are adjusted to minimize
the ground-state energy for an auxiliary Su-Schrieffer-Heeger model.   
(For more details, see Appendix~\ref{appssh}.)
The effective Hubbard repulsion is approximated as
\begin{equation}
\label{ueffdij}
  U_{\rm eff}=U-V_{01}\left\langle{}\frac{d_0}{d_{ij}}\right\rangle_{j(i)},
\end{equation}
with the coefficients
$U=3.63\,t_0$, and $V_{01}=2.03\,t_0$ taken from Ref.\ \cite{Sch13}, and
$\langle{}\dots\rangle_{j(i)}$ denoting the average over three nearest
neighbors $j$ of the site $i$.
(Due to our suppositions on the symmetry, Eq.\ (\ref{ueffdij}) produces
same value for all sites.) 

Depending in the electron-phonon coupling $\beta$, we find the strain of
$\varepsilon_y=0.25$ (for $\beta=2$) or $\varepsilon_y=0.20$ (for $\beta=3$)
is necessary to approach $U_c^{\rm (HF)}$, being a~conventional border of the
CSM range.
This values are comparable with  to the maximal strain of $\approx{}0.20$ 
reported in experiments. The phase of Mott Insulator seems inaccessible
by applying mechanical strains to graphene, although modification of the
equilibrium $U/t_0$ ratio due to substrate effect may possibly enhance the
interaction effects. Below, we discuss the effects of
electron correlations which should be  visible also in CSM (or even SM)
phase.

\begin{figure}[!t]
  \includegraphics[width=0.9\linewidth]{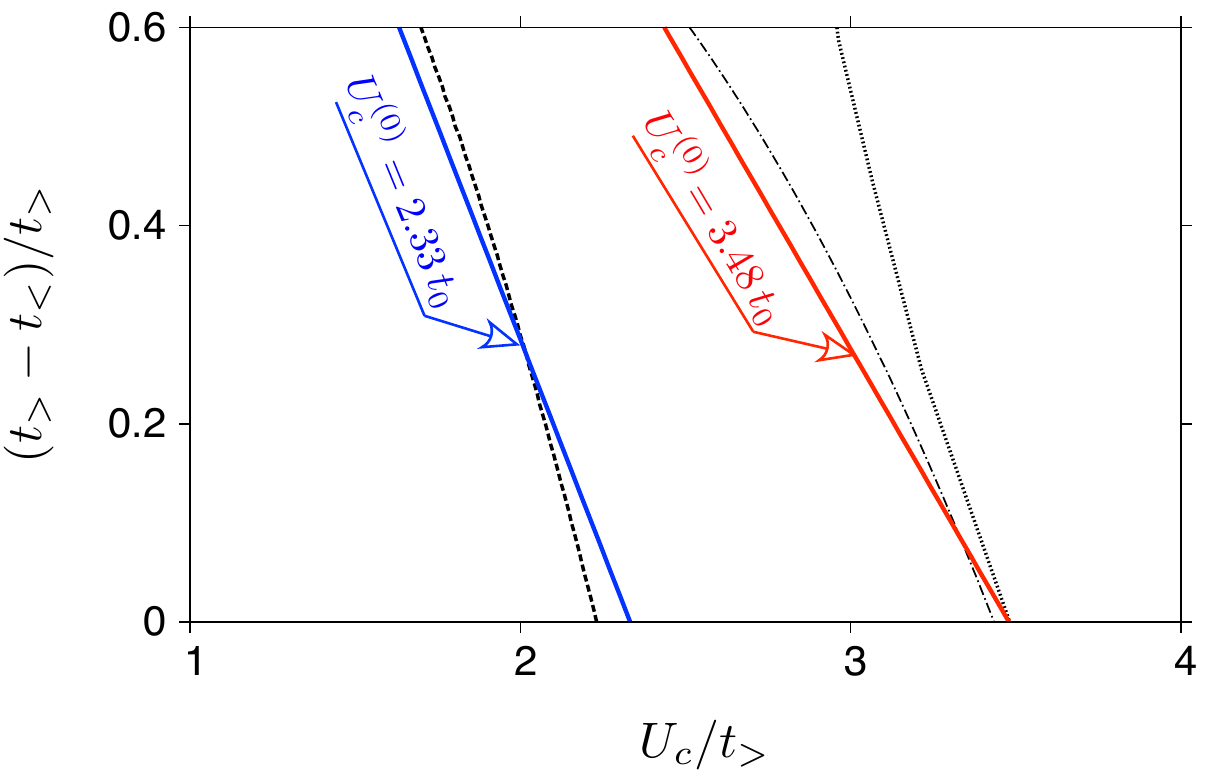}
  \caption{ \label{ucappfig}
    The evolution of critical Hubbard interaction with armchair strain
    strain ($t_y\leqslant{}t_x$), approximated by Eq.\ (\ref{ucapprox})
    [solid lines] for two values of $U_c^{(0)}$ (specified on the plot)
    adjusted to match the zero-strain results obtained from HF and GWF
    methods. Remaining lines are same as in Fig.\ \ref{phadiag2}. 
  }
\end{figure}

\subsection{Effects of strain on measurable quantities}
Earlier in this paper, we point out that a~model assuming linear density
of states, Eq.\ (\ref{toydos}), parametrized by the Fermi velocity a~zero
energy, gives the critical Hubbard interaction $U_c^{(X)}$ that differs by
only $5\%$ from the values obtained using the actual density of states
in the absence of strain, for the two methods, i.e., $X\!=\,$HF and
$X\!=\,$GA. 
It is reasonable to expect, that for strains introducing anisotropy of
the Fermi velocity \cite{Ros12,Ley15} the value of $U_c$ will be affected
predominantly via a~change of the cut-off energy $\Lambda$, related to the
Fermi velocity. For $t_y\leqslant{}t_x$ and strains limited to
experimentally-accessible values, one can set $\Lambda\propto{}\sqrt{t_xt_y}$,
leading to
\begin{equation}
\label{ucapprox}
  U_c\approx{}U_c^{(0)}\left(1-\frac{1}{2}\delta_t\right), \ \ \ \
  \delta_t=\frac{t_x-t_y}{t_x}, 
\end{equation}
where $U_c^{(0)}$ a~zero-strain value. Substituting the values of $U_c^{\rm (HF)}$
and $U_c^{\rm (GWF)}$ for $t_y=t_x$, we find (in Fig.\ \ref{ucappfig}) that
the evolution of $U_c/t_x$ with increasing strain is approximated by
Eq.\ (\ref{ucapprox}) quite well for both HF and GWF methods (similar
agreement is observed for GA results, omitted in Fig.\ \ref{ucappfig}),
but not for CPA, which predicts much weaker effects of strain.

\begin{figure}[!t]
  \includegraphics[width=\linewidth]{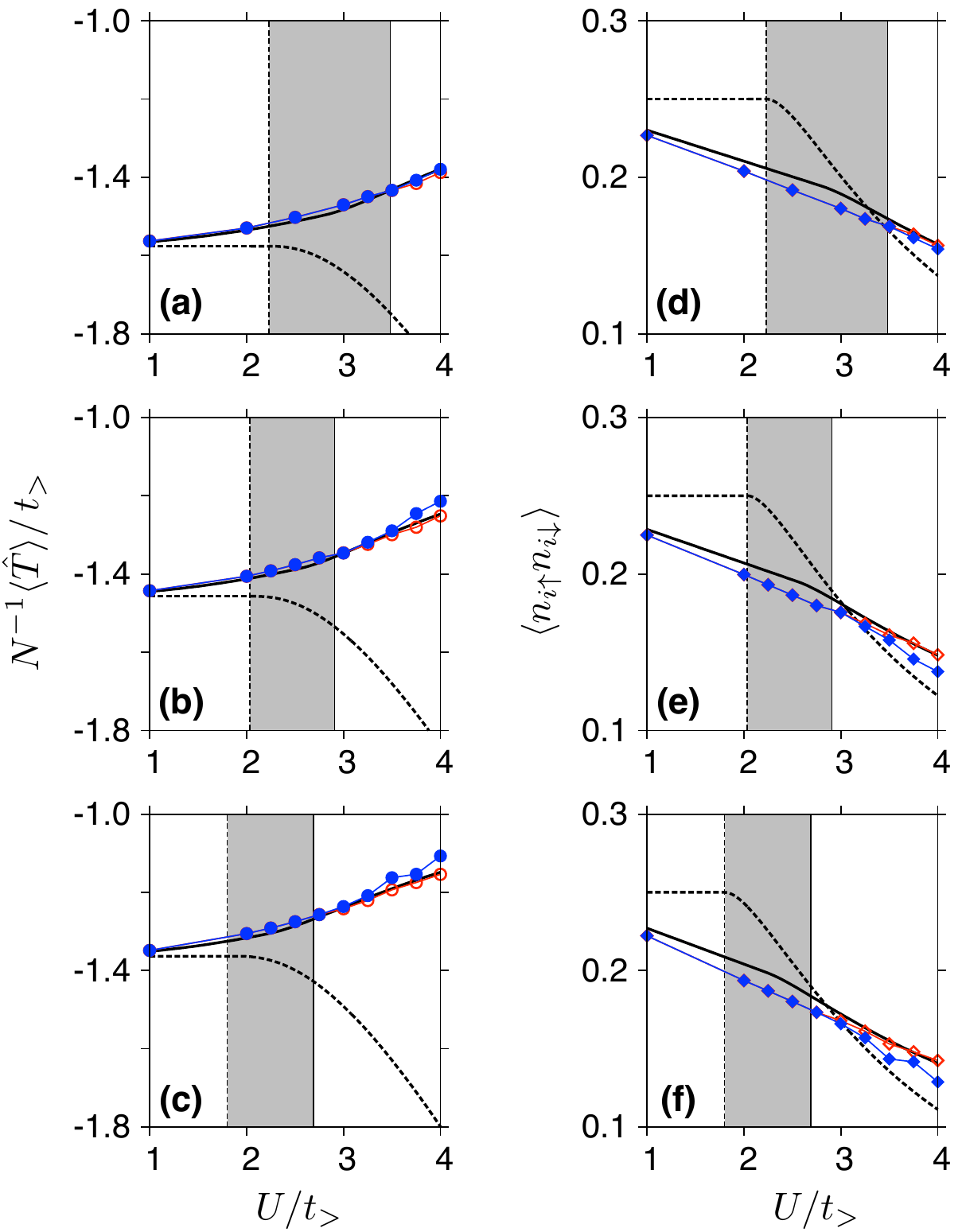}
  \caption{ \label{gwfnund}
    (a)--(c) Average kinetic energy per site and (d)--(f)
    average double occupancy displayed as functions of the Hubbard repulsion
    $U$ for $t_y/t_x=1$ (top), $t_y/t_x=0.75$ (middle),
    and $t_y/t_x=0.5$ (bottom). Thick dashes line marks the Hartree-Fock
    results, thick solid line represents the Gutzwiller Approximation.
    Datapoints depict the VMC results for GWF with $m=0$ (red open symbols)
    and optimized $m$ (blue closed symbols); thin lines are guide for the
    eye only. Shaded area marks the correlated
    semimetallic phase, bounded by $U_c^{\rm (HF)}$ and $U_c^{\rm (GWF)}$. 
    (For the numerical values, see Table~\ref{uctable}.)
  }
\end{figure}

Our results (in particular, a~systematic shrinking of the SM phase, as well
as the CSM range, with increasing strain) suggest that some measurable
signatures of electron correlations should be visible in strained system also
for $U<U_c$.
These expectation is further supported with the data presented in Fig.\
\ref{gwfnund}, where we display the average kinetic energy per site
(quantifying the band narrowing) and the average double occupancy as
functions of $U$.
This time, the GWF results obtained fro VMC simulations do not differer
significantly from GA results (see datapoints and black solid lines,
respectively), while HF (dashed lines) predicts qualitatively different
behavior, particularly for $\langle\hat{T}\rangle$ displayed versus $U$
in Figs.\ \ref{gwfnund}(a), \ref{gwfnund}(b), and \ref{gwfnund}(c),
but the differences between HF and Gutzwiller-based techniques are also
apparent for $\langle{}n_{i\uparrow}n_{i\downarrow}\rangle$ (see remaining panels
in Figs.\ \ref{gwfnund}). 

For $\delta_t=0.5$, see Figs.\ \ref{gwfnund}(c) and \ref{gwfnund}(f),
corresponding to the strain of $\varepsilon_y=0.22$ for $\beta=3$,
and in the interval of $U/t_x=1.5\div{}2$ being relevant for graphene, 
the values of $\langle\hat{T}\rangle$ obtained from GWF or GA are reduced
by more than $20\%$ comparing the $\delta_t=0$ situation; 
see Fig.\ \ref{gwfnund}(a).
(Notice that the above-mention reduction includes the change of $t_x$,
used as an energy unit in Fig.\ \ref{gwfnund}; the details are given
in Appendix~\ref{appssh}.) 
Also in the intermediate case, $\delta_t=0.25$, a~$10\%$ reduction is noticed,
see Fig.\ \ref{gwfnund}(b). 
For $\langle{}n_{i\uparrow}n_{i\downarrow}\rangle$, the effect of strain
is less pronounced, but we still have an approximately $10\%$ reduction for
the $\delta_t=0.5$ case [see Fig.\ \ref{gwfnund}(f)] compared to the
$\delta_t=0$ case [Fig.\ \ref{gwfnund}(d)], following from both GWF or GA
methods for the interval of $U/t_x=1.5\div{}2$.

\section{Concluding remarks}
\label{conclu}

We have investigated the mutual effect of electron-electron interaction,
modeled by a~Hubbard term in the second-quantized Hamiltonian, and geometric
strains applied to a~half-filled honeycomb lattice, quantified (at a~first step)
via two arbitrary values of the nearest-neighbor hopping integrals:
one for bonds inside zigzag lines parallel to a~selected direction, and
the other for remaining bonds.
Related problems were widely studied in the existing literature
\cite{Tan15,Per09,Ros12,Ley15}; therefore, our attention has focussed
on the case when strain is applied in the armchair direction (i.e.,
the hopping integrals connecting different zigzag lines are suppressed). 
In such a~case, energy spectrum for noninteracting system remain gapless
for arbitrary high strains, since the Dirac cones do not merge.
In turn, the semimetal-insulator transition may occur only due to
interactions. Also, the system gradually evolves, with increasing strain,
towards a~collection of weakly-coupled Hubbard chains, allowing to expect
a~considerable reduction of the critical Hubbard repulsion.

Several computational methods are compared, finding that the Hartree-Fock
(HF) approximation, Gutzwiller Approximation (GA), and Gutzwiller Wave Function
(GWF) treated within Variational Monte Carlo simulations, all predict
qualitatively-similar shrinking of the semimetallic phase with increasing
strain. Two remaining methods, the Coherent Potential Approximation (CPA)
and so-called Ne\'{e}l-state GA, produce slightly different shapes of the
semimetal-insulator boundary, but in the CPA case 
the results are numerically close to these obtained from GWF (provided that
the strain is weak or moderate). 

Phase diagram for the parametrized model is supplemented with calculations
of trajectories, followed by monolayer graphene strained in armchair
direction and allowed to relax along the perpendicular (i.e., zigzag)
direction. These calculations were performed employing modified
Su-Schrieffer-Heeger Hamiltonian, including the harmonic terms for bonds and
angles (with the parameters fixed to reproduce elastics properties
for in-plane small deformations), and the term describing coupling between
electrons and the lattice, with dimensionless parameter
varied between the possible values.

Albeit the critical and the actual Hubbard repulsion approach each other
with increasing strain, we find that the semimetal-Mott insulator
transition point cannot be achieved in monolayer graphene subjected to
non-destructive deformations. 
Instead, one can observe the effects of electron correlations, such as
the bandwidth renormalization or the reduction of double occupancies, which
are well-pronounced (and affected by an applied strain) much before
the transition. Probably, the above-mentioned effects will also be relevant
in novel two-dimensional materials, predicted to sustain geometric
deformations up to about 30\% without structural demages \cite{Sin20,Zha22}.
When looking for experimental realization of the Mott insulator on a~honeycomb
lattice, one should rather focus on artificial graphene-like systems
\cite{Sin11,Pol13,Gar20,Tra21}. 

What is more, we have shown that a~basic version of the Gutzwiller
Approximation already captures crucial correlation effects in (strained)
graphene, allowing one to expect that proper generalizations of this method,
such as the diagrammatic expansion \cite{Lid92,Wys14,Fid18b}, may lead to
the development of versatile computational tools for studying graphene and
related Dirac systems, providing low- (or moderate-) costs counterparts for
the Quantum Monte Carlo methods.

\section*{Acknowledgments}
We thank Prof.\ J\'{o}zef Spa\l{}ek for discussions. 
The work was supported by the National Science Centre of Poland (NCN)
via Grant No.\ 2014/14/E/ST3/00256. 
Computations were partly performed using the PL-Grid infrastructure.


\appendix
\setcounter{section}{0}

\section{Coherent potential approximation}
\label{appcpa}

The CPA method \cite{Le13,Row14} employs the alloy-analogue approach,
in which the many-body Hamiltonian (\ref{hamtxyu}), supplemented with
the chemical potential term $-\mu\hat{N}$, is approximated by
\begin{align}
  H-\mu\hat{N} &\stackrel{\rm CPA}{=}
  \sum_{i\in{}A,s}E_{As}n_{is} + \sum_{j\in{}B,s}E_{Bs}n_{is} \nonumber \\
  &+\sum_{\langle{}ij\rangle,s}t_{ij}\left(c_{is}^{\dagger}c_{js}+\mbox{H.c.}\right),
  \label{hamcpa}
\end{align}
where $\hat{N}=\sum_{is}n_{is}$ is the particle number operator, and the
random potential energy
\begin{equation}
  \label{randeas}
  E_{\alpha{}s}=\begin{cases}
  -\mu+U  & \text{with probability }\ n_{\alpha{}\overline{s}}, \\
  -\mu     & \text{with probability }\ 1-n_{\alpha{}\overline{s}}. \\
  \end{cases}
\end{equation}
Here, $\alpha=A,B$ is the sublattice index, $\overline{s}$ denotes the spin
opposite to $s$, and $n_{\alpha{}\overline{s}}$ is the average occupation for
spin $\overline{s}$ in the sublattice $\alpha$.
The half filling corresponds to $\mu=-U/2$.

The Green function for a~single-particle Hamiltonian defined by Eqs.\
(\ref{hamcpa}) and (\ref{randeas}) needs to be averaged over all possible
configurations of the random potential energies.
Within the CPA, energies $E_{\alpha{}s}$ are approximated by self-energies
$\Sigma_{\alpha{}s}(\omega)$, same for all atoms belonging to one sublattice. 
In turn, the Green function can be determined by solving the following
system of self-consistent equations
\begin{align}
  G_{\alpha{}s}(\omega) &=
  \frac{F_{\alpha{}s}(\omega)(1-n_{\alpha{}\overline{s}})}{1+F_{\alpha{}s}(\omega)
  \left[ \Sigma_{\alpha{}s}(\omega) + \mu \right]}
  \nonumber \\
  &+
  \frac{F_{\alpha{}s}(\omega)\,n_{\alpha{}\overline{s}}}{1+F_{\alpha{}s}(\omega)
  \left[ \Sigma_{\alpha{}s}(\omega) + \mu - U\right]},
  \label{gfom1}
  \\
  G_{\alpha{}s}(\omega) &= F_{\alpha{}s}(\omega),
  \label{gfom2}
\end{align}
where $F_{\alpha{}s}(\omega)$ is the local Green function for the sublattice
$\alpha$,
\begin{equation}
  F_{\alpha{}s}(\omega) = \xi_{\overline{\alpha}s}\int{}\frac{dE\,\rho(E)}{
    \xi_{As}\xi_{Bs}-E^2
  }, 
\end{equation}
with $\xi_{\alpha{}s}=\omega+i\eta-\Sigma_{\alpha{}s}(\omega)$ and $\rho(E)$
the density of states defined by Eq.\ (\ref{rhodef}).
Here, $\eta>0$ is a~small real number (typically, we took $\eta/t_x=10^{-3}$).

Eqs.\ (\ref{gfom1}) and (\ref{gfom2}) can be solved iteratively via 
\begin{equation}
  \Sigma_{\alpha{}s}(\omega) = \Sigma_{\alpha{}s}(\omega) +
  \frac{1}{F_{\alpha{}s}(\omega)}-\frac{1}{G_{\alpha{}s}(\omega)}. 
\end{equation}
For simplicity, we limit our discussion to the paramagnetic solution,
$n_{\alpha{}s}=n_{\alpha{}\overline{s}}=1/2$ for $\alpha=A,B$.
The multiparticle density of states is given (up to a~normalization)
by the imaginary part of $-G_{\alpha{}s}(\omega)$. 
This density of states, discussed as a~function of $U$ at a~fixed $\omega=0$,
allows one to estimate the critical Hubbard repulsion $U_c^{\rm (CPA)}$,
corresponding to the semimetal-insulator transition. 
Namely, $U_c^{\rm (CPA)}$ can be identified as the point when a~nonzero
$-\mbox{Im}\,G_{\alpha,s}(0)$ rapidly drops to zero with increasing $U$;
see Ref.\ \cite{Le13}.

\section{Su-Schrieffer-Heeger model for graphene}
\label{appssh}

\begin{figure}[!t]
  \includegraphics[width=\linewidth]{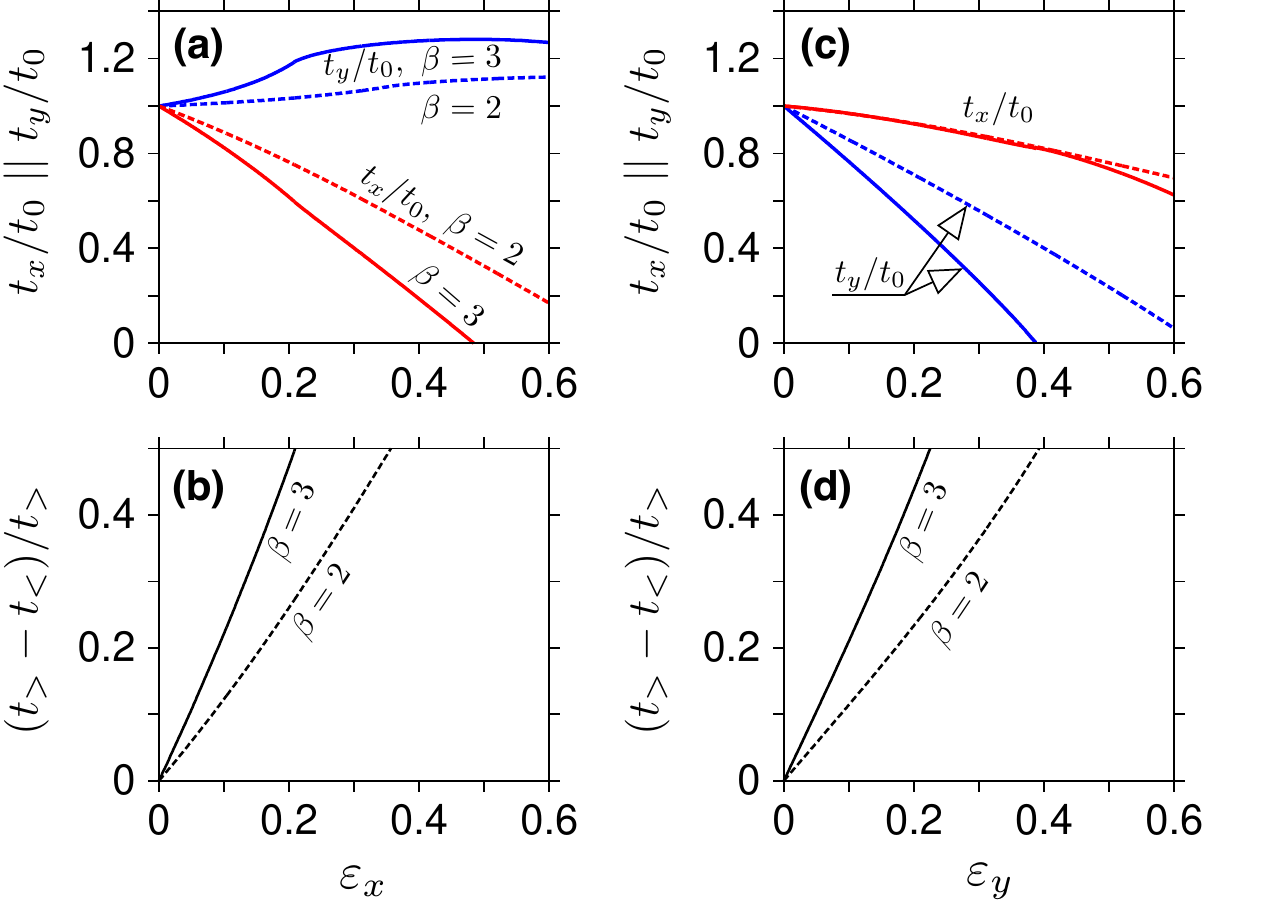}
  \caption{ \label{txyfig}
    The hopping-matrix elements [see Eqs.\ (\ref{tijtxy}) and (\ref{tijdij})
    in the main text] as functions of the strain $\varepsilon_x>0$ (a), (b)
    or $\varepsilon_y>0$ (c), (d). The electron-phonon coupling is fixed
    at $\beta=2$ (dashed lines in all plots) or $\beta=3$ (solid lines). 
  }
\end{figure}

Below, we put forward a~modified Su-Schrieffer-Heeger (SSH) model for
tight-binding electrons adiabatically coupled to acoustic phonons
\cite{Dre98,Gro18,Ryc13}, allowing one to map physical strains,
applied to graphene, onto the Hamiltonian (\ref{hamtxyu}).
For this purpose, we consider 
\begin{align}
  H_{\rm SSH} &= -t_0 \sum_{\langle{}ij\rangle,s}
    \left({1-\beta\frac{\delta{}d_{ij}}{d_0}}\right)
  \left(
    c_{i,s}^\dagger{}c_{j,s}+\mbox{H.c.}
  \right) \nonumber \\
  &+
  \frac{1}{2}K_d\sum_{\langle{}ij\rangle}
  \left(d_{ij}\!-\!\tilde{d}_0\right)^2
  +
  \frac{1}{2}K_{\theta}\!\sum_{j,{\measuredangle(j)}}\!
  \left(\theta_{\measuredangle(j)}\!-\!\theta_0\right)^2.
  \label{hamssh}
\end{align}
Here, $t_0=2.7\,$eV is the equilibrium nearest-neighbor hopping integral
for $\pi$ electrons in monolayer graphene,
$\beta=-\left.\partial{}\ln{}t_{ij}/\partial\ln{}d_{ij}\right|_{d_{ij}=d_0}$
is the dimensionless parameter quantifying the electron-phonon coupling
(later, we perform main calculations for $\beta=2$ and $\beta=3$), and
$\delta{}d_{ij}=d_{ij}-d_0$ is the bond-length change, calculated with respect
to the equilibrium length of $d_0=a/\sqrt{3}=0.142\,$nm.
The second and third term in $H_{\rm SSH}$ represent potential energy
describing the covalent bonds \cite{Tsa10}, with $\measuredangle(j)$
denoting the angles having a~common vertex at a~given lattice site $j$
(see Fig.\ \ref{setupxy}). 

The parameters $K_d=40.67\,$eV$/$\AA$^2$, $K_\theta=5.46\,$eV$/$rad$^2$,
and $\theta_0=\pi/3$, are adjusted to restore the in-plane elastic
coefficients of bulk graphene in the $\beta=0$ case \cite{Tsa10}.
For $\beta\neq{}0$, a~correction to the effective potential energy per bond
$(2/3)\left.\partial^2\epsilon_0/\partial{}d_{ij}^2\right|_{\{d_{ij}=d_0\}}=0$, 
provided that the equilibrium bond length is unaffected.
To guarantee the last condition, we introduce the effective
($\beta$-dependent) equilibrium length
\begin{equation}
  \tilde{d}_0(\beta) = d_0+\frac{2\beta{}\epsilon_0}{3K_dd_0}
  \approx{}d_0(1+0.03455\,\beta), 
\end{equation}
where we have substituted the equilibrium kinetic energy per site
$\epsilon_0=-1.57460\,t_0$. (Notice that a~standard constrain to the SSH
model, $\sum_{\langle{}ij\rangle}d_{ij}=\,$const., is irrelevant when studying
graphene with a~global strain.)

Next step is the optimization of the ground-state energy,
\begin{equation}
  \label{egssh}
  E_G^{\rm (SSH)}(\{{\bf R}_j\})=\langle{\cal H}_{\rm SSH}\rangle, 
\end{equation}
with  respect to in-plane atomic positions $\{{\bf R}_{j}\}$.
In this paper, we limit the discussion to atomic arrangements preserving
the bipartite structure structure of the lattice and the two mirror
symmetries. For a~fixed strain in the selected direction, 
$\varepsilon_>=\max(\varepsilon_x,\varepsilon_y)$, there are two parameters
left to be optimized: the elongation in the perpendicular direction,
$\varepsilon_<=\min(\varepsilon_x,\varepsilon_y)$, 
and the length $d_y$ of the bonds parallel to $y$-axis.
The length of the remaining bonds, belonging to zigzag line,
is given by
\begin{equation}
  d_x = d_0\left[\frac{3}{4}\left(1+\varepsilon_x\right)^2
  +\frac{9}{4}\left(1+\varepsilon_y-\frac{2d_y}{3d_0}\right)^2\right]^{1/2}. 
\end{equation}
Numerical values of the hopping matrix elements, following from the
optimization procedure for $\beta=2$, $3$, and the two directions of strain
are displayed in Fig.\ \ref{txyfig}.

In principle, the optimization scheme similar to the above can be
constructed also for the Hamiltonian containing both lattice degrees of
freedom and the Hubbard repulsion, $H_{\rm SSH}+U\hat{D}$.
For instance, if $U<U_c$ and the paramagnetic phase can be assumed, one
can refer directly to Eq.\ (\ref{eggam0}), and setup a~self-consistent
procedure sharing main features of the EDABI method \cite{Spa07}, but
not limited to small systems.
However, the term $\propto{}U^2$ in Eq.\ (\ref{eggam0}), introducing 
the coupling between lattice and electron correlations, is preceded
by a~small prefactor and thus the corrections to atomic arrangements
due to $U>0$ are insignificant (notice that the Hubbard repulsion for
a~monolayer in equilibrium is $U_{\rm eff}\approx{}1.6\,t_0$ \cite{Sch13}). 
For these reasons, we decided not to pursue this direction here, 
limiting our presentation to the data obtained
directly via minimizing $E_G^{\rm (SSH)}(\{{\bf R}_j\})$ in Eq.\ (\ref{egssh}),
leaving the geometric parameters and the correlation effects decoupled.




\begin{thebibliography}{}
\bibitem{kxyfoo}
Due to symmetry, it sufficient to sum over first quarter of the hexagonal
Brilloun zone, namely, for $0\leqslant{}k_y<2\pi/\sqrt{3}$ and 
$0\leqslant{}k_x<4\pi/3 - |k_y|/\sqrt{3}$.
\bibitem{vmcfoo}
Typically, for each combination of $t_y/t_x$, $U/t_x$ and $\Delta/t_x$,
we took $15\div{}20$
values of the parameter $g=e^{-\eta}$ separated by the steps of $0.01$,
in the vicinity of a~predicted energy minimum. 
For each $g$, the averages over distributions of electrons in real space
were calculated by performing $10^5$ iterations per lattice, according to
the Glauber's algorithm.
Initial $10^4$ iterations per site was neglected for each simulation, to
avoid the effects of initial configuration.
The variational energy $E_G^{\rm (GWF)}$ (and corresponding optimal value of
the parameter $g$) was then determined via the least-squares fitting of
a~quadratic function.
\end{thebibliography}
\end{document}